\documentclass{article}

\pdfoutput=1

\usepackage{epsfig}
\usepackage{graphics}
\usepackage{graphicx}
\usepackage{amsmath}
\usepackage{amsfonts}
\usepackage{amsthm,subfigure,url}
\usepackage{amssymb}
\usepackage{url,subfigure}
\usepackage{rotating}

\begin{document}


\title{Hot Ice Computer}

\author{Andrew Adamatzky}

\markboth{Adamatzky}{Hot Ice Computer}

\date{}

\maketitle

\vspace{0.5cm}

\begin{center}
 Department of Computer Science, University of the West of England,\\ Bristol, United Kingdom \\
\url{andrew.adamatzky@uwe.ac.uk}
\end{center}

\vspace{0.5cm}

\begin{abstract}
\noindent

We experimentally demonstrate that supersaturated solution of sodium acetate, commonly called `hot ice', 
is a massively-parallel unconventional computer. In the hot ice computer data are represented 
by a spatial configuration of crystallization induction sites and physical obstacles immersed
in the experimental container. Computation is implemented by propagation and interaction of growing 
crystals initiated at the data-sites. We discuss experimental prototypes of hot ice processors 
which compute planar Voronoi diagram, shortest collision-free paths and implement {\sc and} and {\sc or}
logical gates.   

\vspace{0.5cm}

\noindent
\textbf{Keywords:} crystallization, Voronoi diagram, shortest path, unconventional computing, 
logical gates, physics of computation 
\end{abstract}

\section{Introduction}

The last decade has witnessed a rise in the number of unconventional computing devices based on principles and mechanisms 
of information processing in physical, chemical and biological systems~\cite{uc_2007}.  Most results 
in unconventional computing are theoretical~\cite{uc08,uc09} whilst there is a handful of intriguing 
experimental prototypes. The experimental laboratory research focuses on implementation of 
computing schemes in novel materials which helps us to grasp the inner nature of physics of computation. 
The following prototypes, or their families, give us a good representation of novel computing substrates. 

Reaction-diffusion computers are implemented in a spatially extended chemical media. Data and results of computation
are represented by spatial distributions of reagent concentrations. Computation is realized by propagating and interacting 
diffusive or excitation waves~\cite{rdc}. Experimental laboratory prototypes of reaction-diffusion computers solve a wide range of geometrical problems, optimization  and logical computation~\cite{rdc}.  

Physarum computing is based on adaptive foraging behaviour of plasmodium of \emph{Physarum polycephalum}.
In \emph{Physarum} computers data  are represented by the distribution of attracting (food) 
and repelling sources (light). Computation is implemented by the plasmodium which optimizes it feeding 
pattern under control of  attracting/repelling forces. Results of the computation are represented by 
the morphological structure of the plasmodium~\cite{nakagaki_2001a,nakagaki_iima_2007,adamatzky_ppl_2007,adamatzky-bz-trees}.

Extended analog computers represent data as a spatial configuration of point-wise sources of 
current applied to a conductive medium. Computing operations are determined by 
physical properties of the conductive medium, (e.g. foam or gel). Results are represented by 
the distribution of voltages across the medium~\cite{mills}.

Other less-known prototypes include analog computing schemes using microwaves~\cite{everitt},  
plane tessellation in gas-discharge semi-conductor systems~\cite{astrov}, learning networks 
of memristors~\cite{erokhin}, logical circuits in liquid crystals~\cite{harding}, and 
glow discharge systems computing the shortest path~\cite{reyes}.

Most experimental prototypes of unconventional computers either require a tailored hardware interface 
(analog computers, liquid crystals) and specialized equipment (memristors, gas-discharge systems), or they may have intrinsic limitations on the speed of computation (reaction-diffusion chemical processors). \emph{Physarum} computer
is the simplest to build but the most difficult to control, due to the sensitivity and somewhat unpredictable behaviour of the living creature.

We aim, therefore, to provide an example of a novel computing material which is cheap to build, 
requires minimal resources to operate, implements computational procedures relatively 
quickly and is capable of solving a wide range of computationally-hard tasks. We show that 
sodium acetate trihydrate (colloquially called `hot ice' due to its resemblance to ice and its crystalline behaviour) perfectly fits our specification 
of an `ideal DIY unconventional computer' because it solves a variety of tasks by traveling 
and interacting waves of crystallization in its supersaturated solution.   
 
The paper is structured as follows. Experimental methods are outlined in Sect.~\ref{methods}. Section~\ref{voronoi}
presents experimental results on computing the Voronoi diagram.  In Sect.~\ref{detecting} we show how to extract 
direction towards the site of crystallization induction. Computation of one-destination-many-sources paths by crystallization 
of sodium acetate, and further extraction of a collision-free shortest path,  are discussed in Sect.~\ref{shortestpath}.
Experimental implementation of logical gates is presented in Sect.~\ref{gates}. We summarize the findings and discuss 
further developments in Sect.~\ref{discussion}.

\section{Methods}
\label{methods}

We prepare a supersaturated clear solution of sodium acetate trihydrate \linebreak CH$_3$COONa$\cdot$3H$_2$O, pour the 
hot solution into Petri dishes and cool the solution down to -5$^o$C. To induce crystallization we briefly immerse 
aluminum wire (powdered with fine crystals of the sodium acetate) into the solution.  

To compute shortest paths in a space with obstacles we mimic obstacles with blobs of silicon, small Petri dishes 
fixed to the bottom of experimental container and labyrinths made of Blu-Tack$^{\textregistered}$\footnote{Bostik Ltd., \url{http://www.bostik.co.uk/}}. 

Dynamics of crystallization is recorded with FujiPix 6000 digital camera and 
still high-resolution images are produced by scanning containers with crystallized sodium acetate in HP Deskjet 5100 scanner. 
Magnified images of the crystalline structure formed are taken using Digital Blue$^{\textregistered}$ QX-5 USB microscope.

Software tools for extracting directions of crystal growth from still images and models of cellular-automaton shortest path 
calculation are coded in Processing\footnote{\url{www.processing.org}}.  

\section{Voronoi diagram}
\label{voronoi}

Let $\bf P$ be a nonempty finite set of planar points. A planar Voronoi diagram of $\bf P$ 
is a partition of the plane into such regions, that for any element of $\bf P$, a region 
corresponding to a unique point $p$ contains all those points of the plane which are 
closer to $p$ than to any other node of $\bf P$. A unique region 
$vor(p) = \{z \in {\bf R}^2: d(p,z) < d(p,m) \forall m \in {\bf R}^2, \, m \ne z \}$ assigned 
to point $p$ is called a Voronoi cell of the point $p$. The boundary of the Voronoi cell of 
a point $p$ is built of segments of bisectors separating pairs of geographically closest 
points of the given planar set $\bf P$. A union of all boundaries of the Voronoi cells 
determines the planar Voronoi diagram:
$VD({\bf P}) = \cup _{p \in {\bf P}} \partial vor(p)$~\cite{shamos_preparata,klein_1990}. 

A basic concept of parallel approximation of Voronoi diagram by wave patterns propagating in a 
spatially extended medium~\cite{adamatzky_1994} is based on time-to-distance transformation. 
To compute a bisector separating two given points $p$ and $q$ we initiate wave-patterns in 
the medium's sites geographically corresponding to $p$ and $q$. The waves travel the same 
distance from the sites of origination before colliding. The loci where the waves meet indicate 
sites of the computed bisector. 

Precipitating reaction-diffusion chemical media are proved to be an ideal computing substrate for approximation of the
planar Voronoi diagram~\cite{tolmachiev,ben_2004,rdc}. A Voronoi diagram can be approximated in a two-reagent medium. 
One reagent $\alpha$  is saturated in the substrate, drops of another reagent $\beta$ are applied to the sites corresponding to planar points to be separated by bisectors. The reagent $\beta$ diffuses in the substrate and reacts with reagent $\alpha$. Colored precipitate is produced in the reaction between $\alpha$ and $\beta$. When two or more waves of diffusing $\alpha$ meet, no precipitate is formed~\cite{ben_2004}. Thus uncolored loci of the reaction-diffusion medium represent bisectors of the computed diagram.  

Most reaction-diffusion chemical processors are quite slow in computing a Voronoi diagram at the macro-scale, due to speed 
limitations caused by diffusion. Crystallization patterns in sodium acetate propagate much quicker and thus could 
improve speed limitations of existing reaction-diffusion algorithms. 

\begin{figure}
\centering
\subfigure[]{\includegraphics[width=0.49\textwidth]{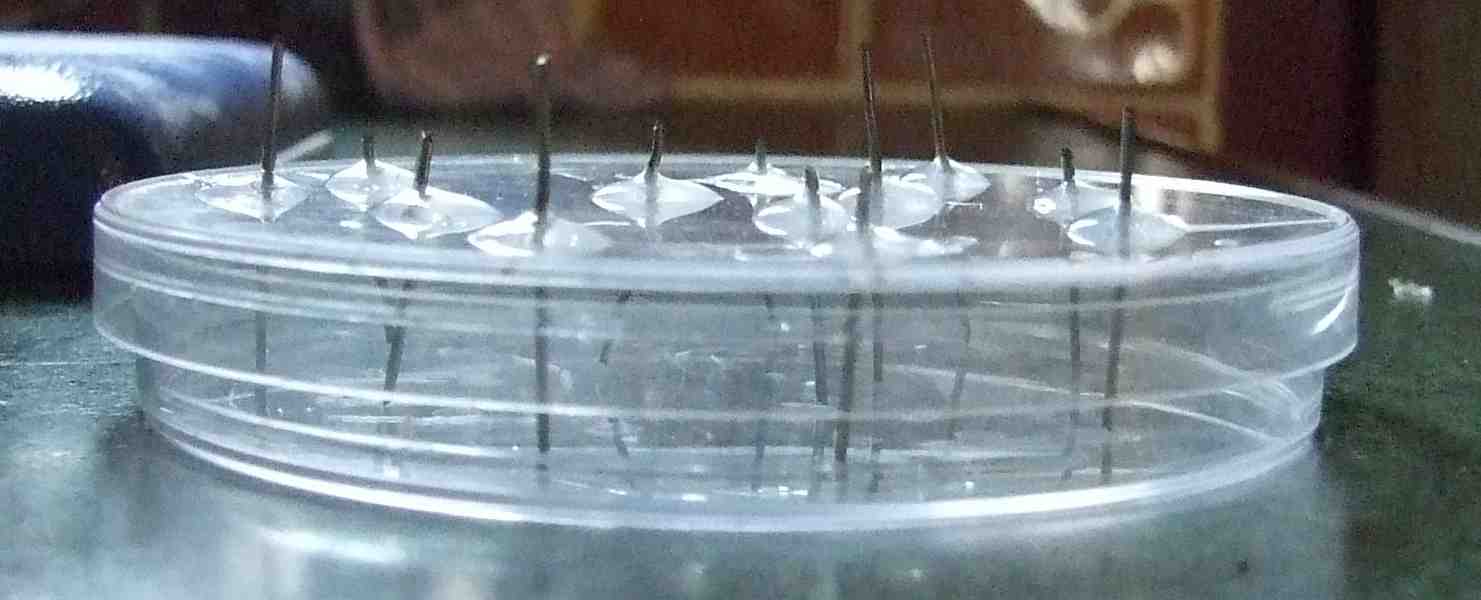}}
\subfigure[]{\includegraphics[width=0.49\textwidth]{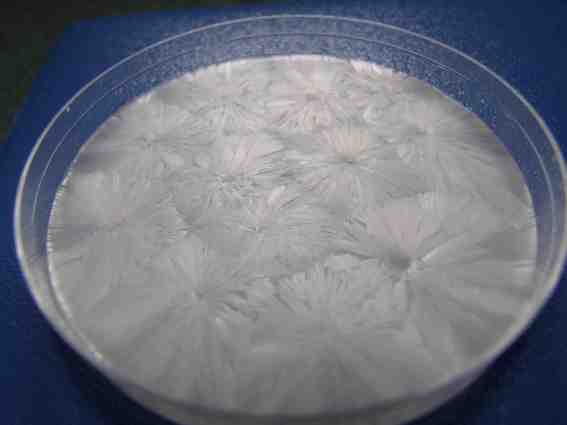}}
\subfigure[]{\includegraphics[width=0.49\textwidth]{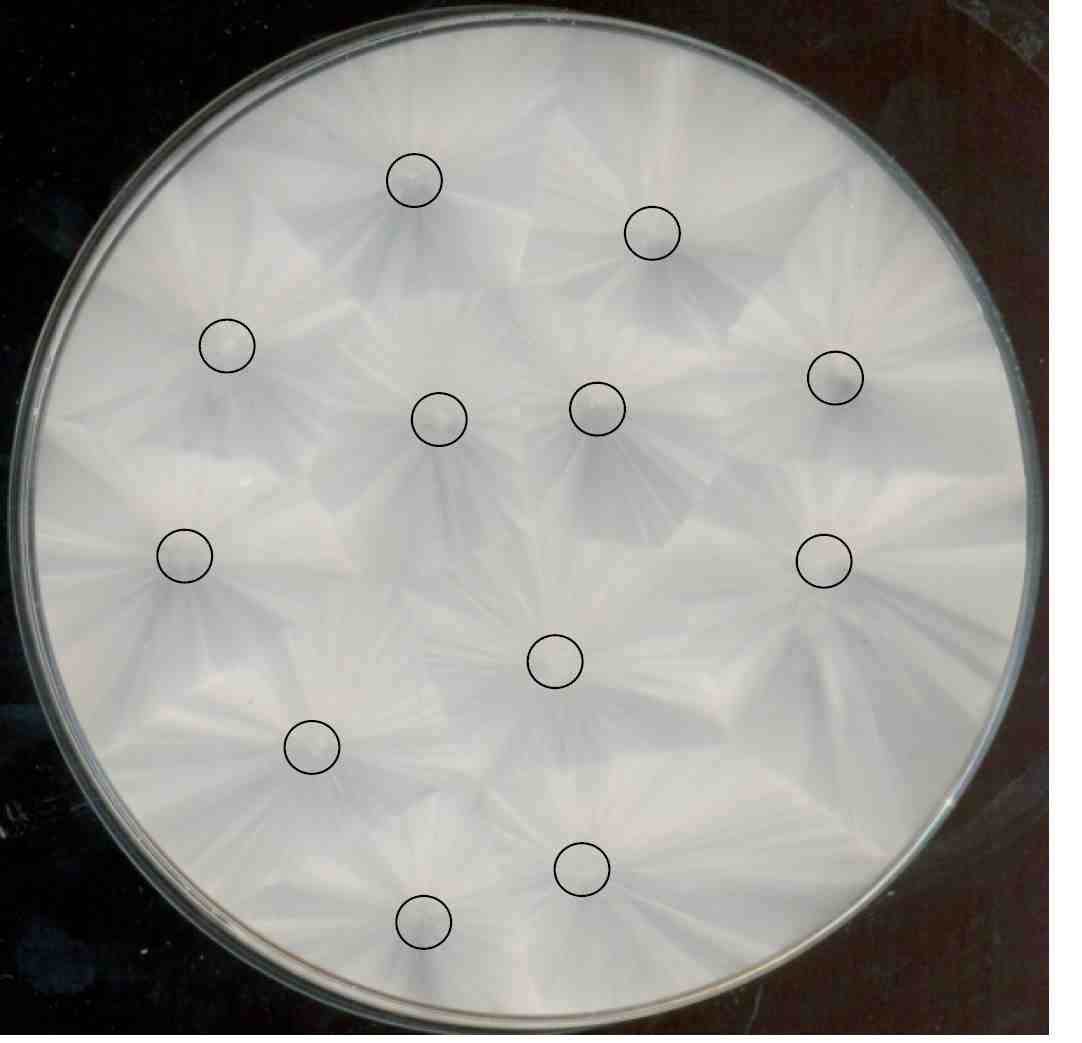}}
\subfigure[]{\includegraphics[width=0.49\textwidth]{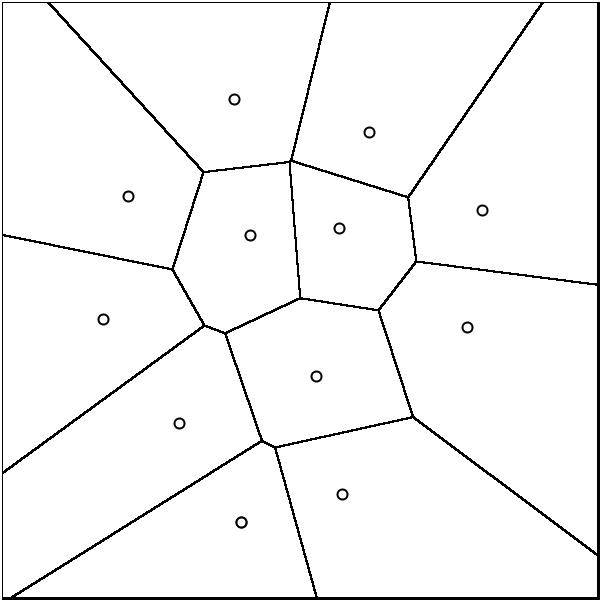}}
\caption{Computation of Voronoi diagram by crystallization of sodium acetate: 
(a)~configuration of pins representing data points $\bf P$, 
(b)~photograph and (c)~scanned image (sites where crystallization was induced are encircled) 
of crystal structure induced by the pins,
(d)~planar Voronoi diagram of $\bf P$ computed by classical algorithm. See video of the experiment at \cite{demos}.}
\label{vd_experimental}
\end{figure}

We represented data points of set $\bf P$ by a configuration of pins (pieces of aluminium wire) 
fixed through the lid of a 9~cm Petri dish 
(Fig.~\ref{vd_experimental}a). The pins were powdered with fine crystals of sodium acetate. 

To start computation we place the lid on the dish with the supersaturated solution. The pins become immersed into the solution.
They induce crystallization. Patterns of crystallization propagate -- as classical target waves -- from the 
sites of crystallization induction. 

A crystallization wave stops when it encounters another wave of crystallization 
(Fig.~\ref{vd_experimental}bc). Boundaries between stationary patterns of crystallization 
represent edges of planar Voronoi diagram  $VD({\bf P})$. The experimental boundaries 
perfectly correspond to bisectors of the diagram computed by the classical Fortune's sweepline 
algorithm~\cite{fortune_1986} (Fig.~\ref{vd_experimental}d).

\section{Detecting directions towards sites of crystallization induction}
\label{detecting}

\begin{figure}
\centering
\subfigure[]{\includegraphics[width=0.49\textwidth]{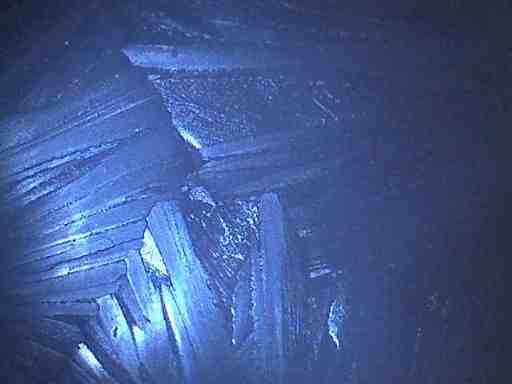}}
\subfigure[]{\includegraphics[width=0.49\textwidth]{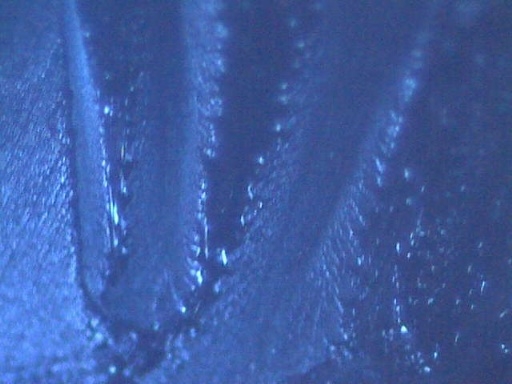}}
\subfigure[]{\includegraphics[width=0.49\textwidth]{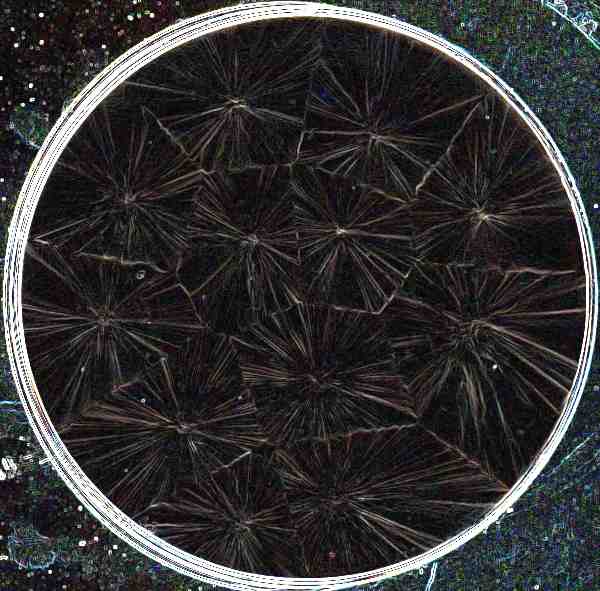}}
\subfigure[]{\includegraphics[width=0.49\textwidth]{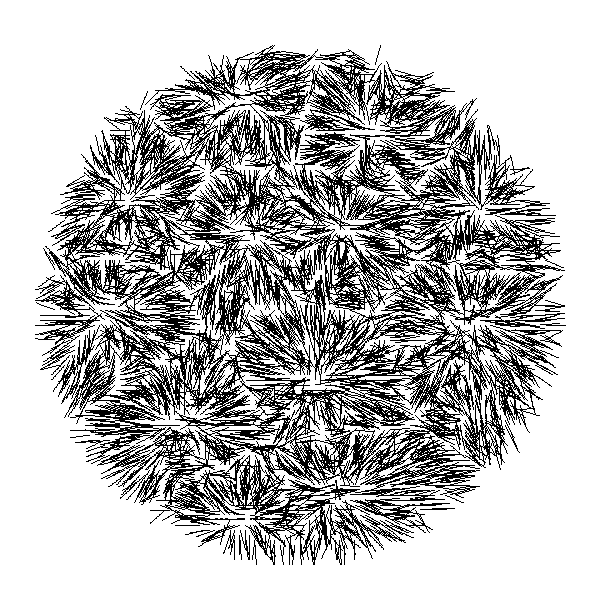}}
\caption{Detecting directions towards sites of crystallization induction:
crystal pattern in $\times$10~(a) and $\times$60~(b) magnification, 
(c)~edge detection operation is applied to image Fig.~\ref{vd_experimental}c, 
(d)~sample configuration of local direction vectors computed from image 
Fig.~\ref{vd_experimental}c.}
\label{crystals}
\end{figure}

Crystals growing from from nucleation sites bear distinctive elongated shapes expanding towards 
their proximal ends (Fig.~\ref{crystals}a). Not only a crystal's overall shape but also the orientation of 
saw-tooth edges indicate the direction of the crystal's growth (Fig.~\ref{crystals}b).   

The direction of crystal growth can be detected by conventional 
image processing techniques, e.g. edge detection procedures (Fig.~\ref{crystals}), or 
by a complementary method of detecting directional uniformity of image domains as 
discussed below. 

We detect local directions of crystal grows in the following way. A fine grid of nodes is applied onto
given image and for every node $x$ of the grid we calculate a set $\bf V$ of vectors of length $r$ 
originating at $x$ and directed at angles $\alpha \in \{ 0, 0.1, 0.2, \cdots, 2/\pi \}$. For 
each vector $v \in \bf V$ we calculate a sum of standard deviations $\sigma(v)$ of RGB colors of 
the image pixels coinciding with $v$. A vector $v' = \min_{v \in \bf V} \sigma(v)$  with minimal 
sum of color deviations indicates  a local direction of crystallization propagation. 
 A configuration of local direction vectors, $r=20$, is shown in Fig.~\ref{crystals}d. 

The vector configuration represents sinks determined by points from set $\bf P$. If we place a
mobile agent at any site but bisectors of the vector configuration Fig.~\ref{crystals}d the agent will be attracted
to the closest data point $p$. This leads us to another problem solvable with sodium acetate --- computation of a collision-free 
shortest path in a space with obstacles.

\section{Computation of a collision-free path}
\label{shortestpath}

 Previously~\cite{adamatzky_1996} we have designed a cellular-automaton algorithm for constructing one-destination-many-sources 
 shortest path. The automaton is a regular array of finite state machines (cells) which takes discrete states and update their 
 states simultaneously and in discrete time depending on the states of its closest neighbors. Here we consider hexagonal array of 
 cells.  Each cell has six neighbors and takes three states --- resting, excited and refractory. A resting cell becomes excited
 if it has at least one excited neighbour. An excited cell switches to a refractory state, and refractory cell returns to resting   
 state independently on states of its neighbors. When we excite one cell of a resting array a wave of  excitation 
 is initiated. The excitation wave propagates in the array.  
 
\begin{figure}
\centering
\subfigure[]{\includegraphics[width=0.15\textwidth]{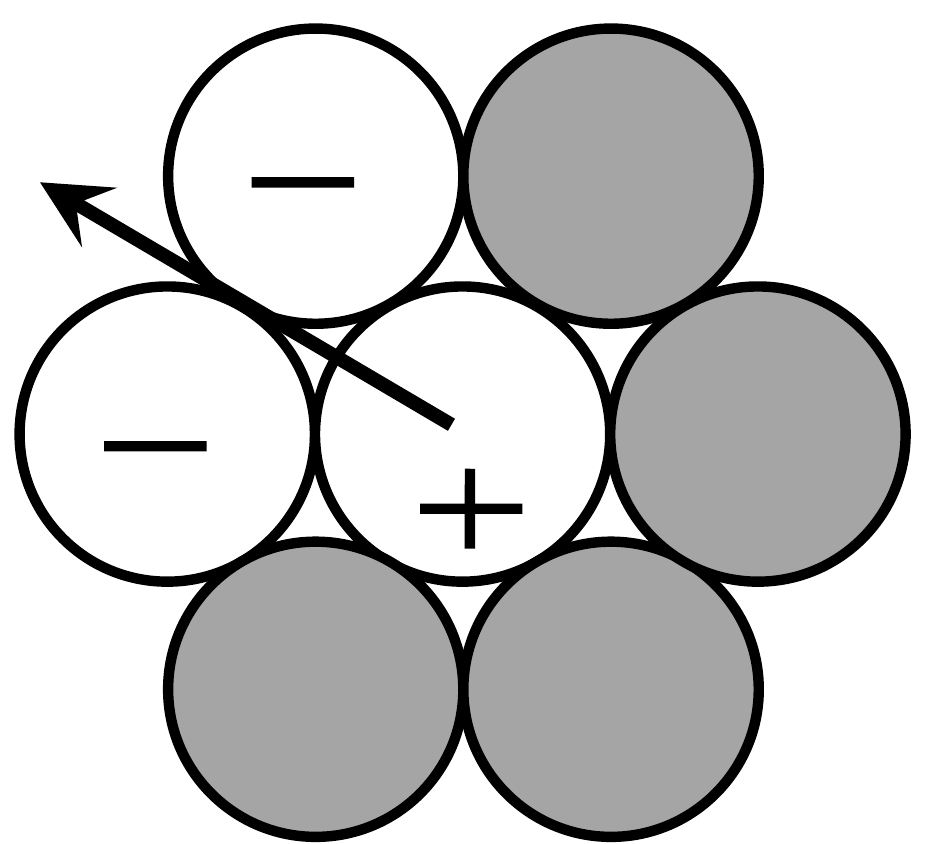}}
\subfigure[]{\includegraphics[width=0.84\textwidth]{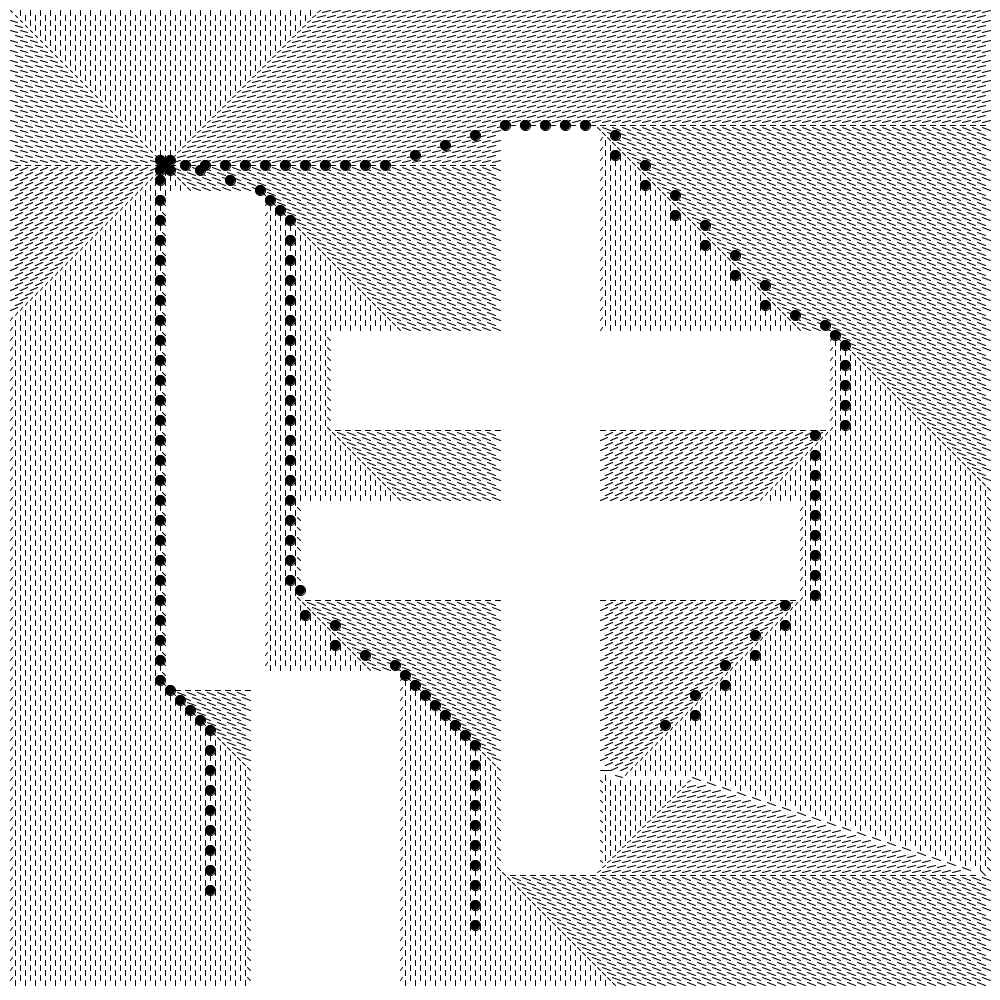}}
\caption{Computation of shortest paths to a single destination: 
(a)~example of pointer orientation (resting cells are gray, excited cell is marked with '+' and refractory cells with '-'),
(b)~example of pointer configuration in presence of obstacles and three shortest paths (bullets) to the site of 
initial excitation; orientation of cell pointers is shown explicitly. }
\label{pathCA}
\end{figure}

 To store the computed path we supply every cell with a pointer. Pointers in all cells are 
 set to nil initially. When a cell becomes excited  its pointer orients towards the direction 
 from where the excitation wave came. An example is shown in Fig.~\ref{pathCA}a.

Obstacles are imitated by always-resting, or non-excitable, cells. 
Examples of the pointer configuration and shortest collision-free paths 
are shown in Fig.~\ref{pathCA}b. A cell assigned to be a destination (in the north-west 
corner of the array Fig.~\ref{pathCA}b) is excited. Waves of excitation propagate 
from the site of initial stimulation. The waves spread around obstacles and disappear in 
the absorbing boundaries of the array. The excitation waves induce orientation of 
cell pointers (Fig.~\ref{pathCA}b).  Array of pointers represents a set of shortest collision-free paths from any cell (source) of the array to the site of initial excitation (destination). Three such paths are shown in Fig.~\ref{pathCA}b.

As we demonstrated in Sect.~\ref{detecting} longitudinal crystals growing in sodium acetate are physical analogs of pointers indicating from where --- locally -- the wave of crystallization came from. That is crystallization of sodium acetate is the 
physical implementation of the original cellular-automaton algorithm~\cite{adamatzky_1996} for computation of shortest 
collision-free path.

\begin{figure}
\centering
\subfigure[]{\includegraphics[width=0.49\textwidth]{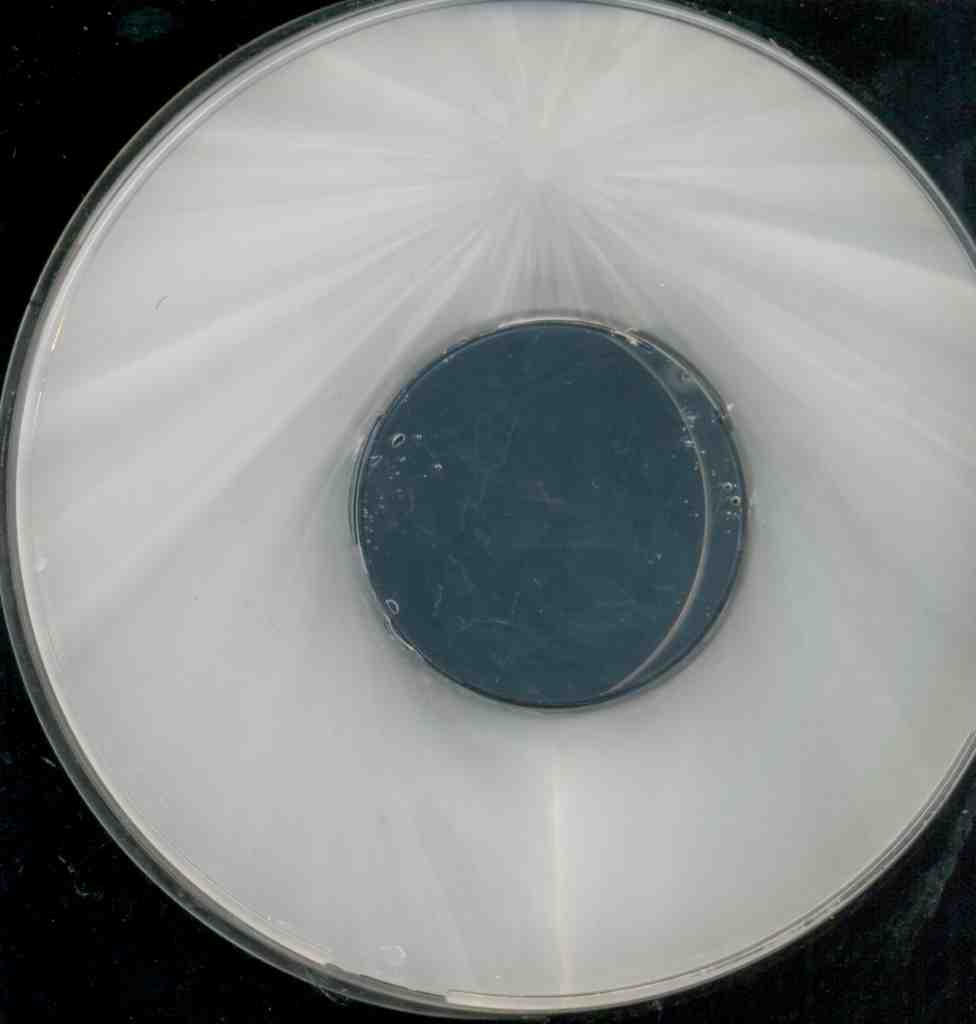}}
\subfigure[]{\includegraphics[width=0.49\textwidth]{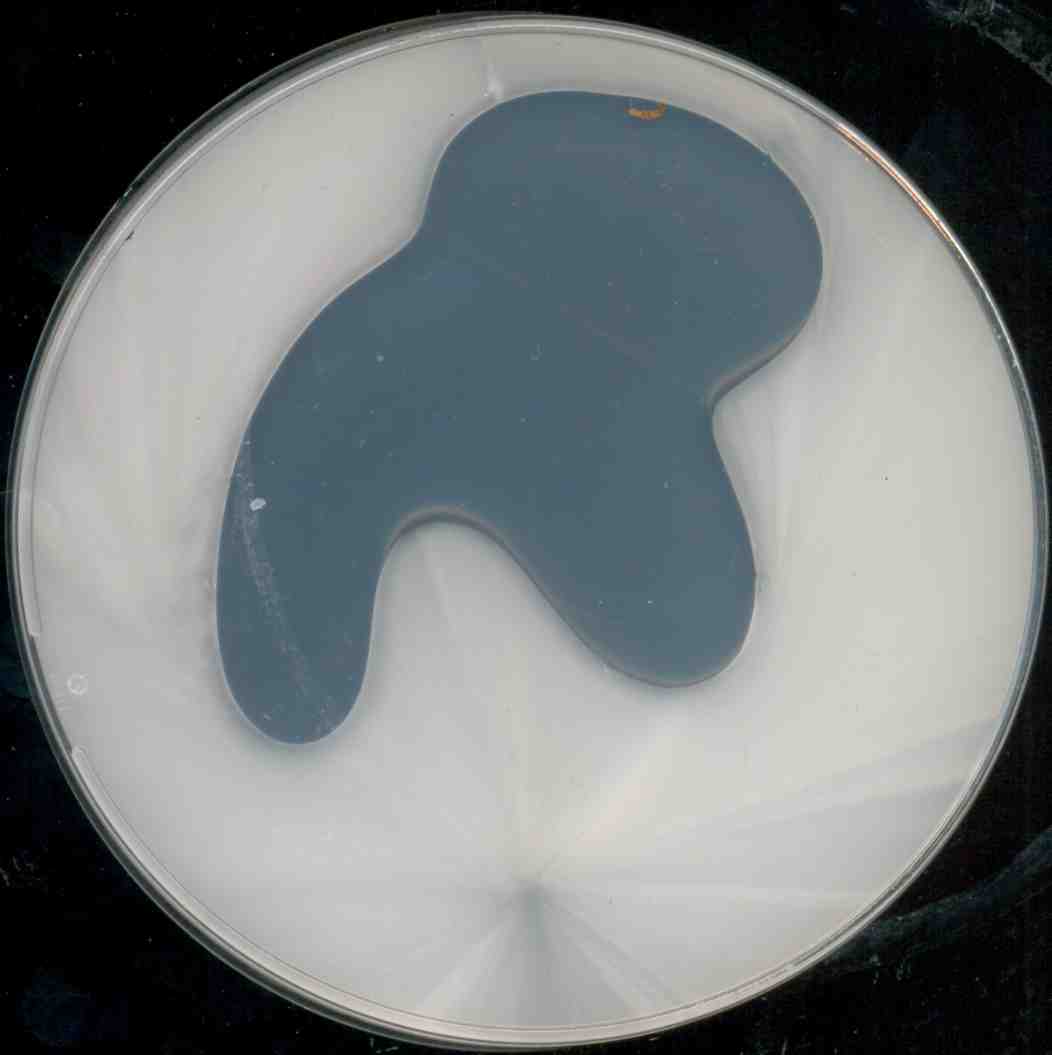}}
\subfigure[]{\includegraphics[width=0.49\textwidth]{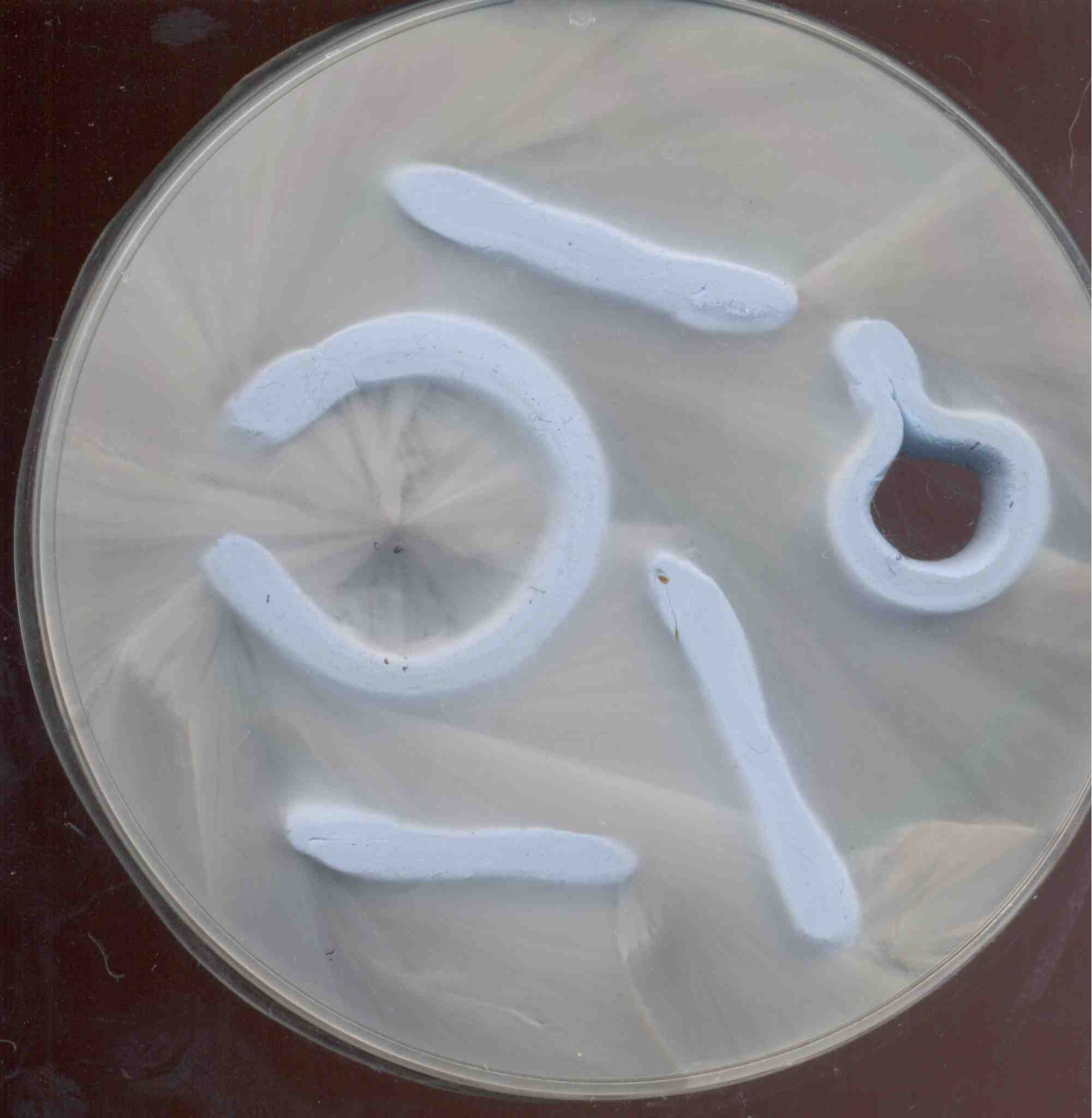}}
\subfigure[]{\includegraphics[width=0.49\textwidth]{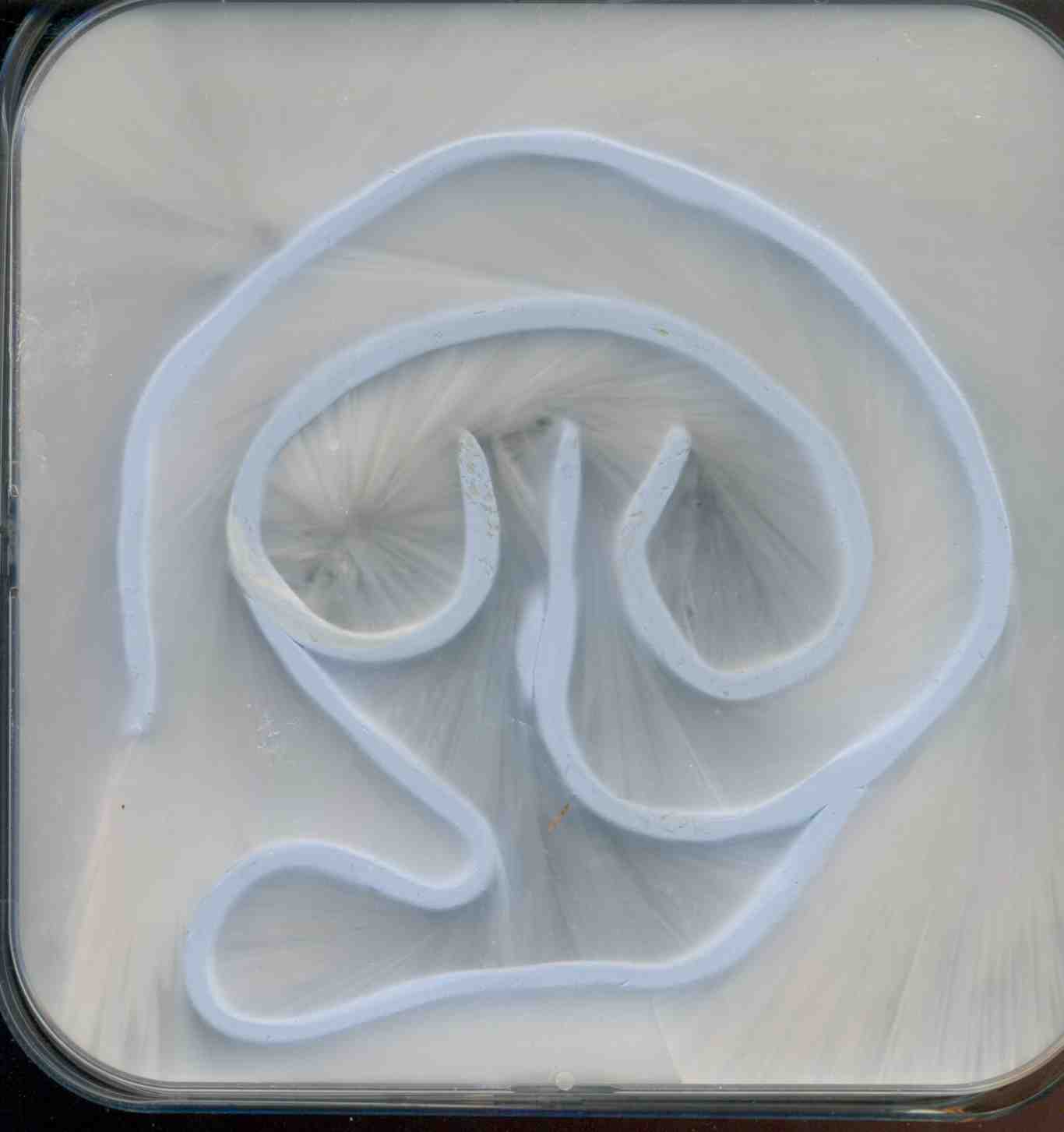}}
\caption{Scanned images of crystallization of sodium acetate occurred in shallow dishes with 
impenetrable obstacles: (a)~35~mm Petri dish glued to the bottom of the container, 
(b)~blob of silicon, and complex obstacles~(c) and maze~(d) made of Blu-Tack$^{\textregistered}$. 
Sites of crystallization induction (destinations of shortest paths) are encircled. See videos 
of the experiments at~\cite{demos}.}
\label{expscans}
\end{figure} 

Let us consider experimental realization.  We represent obstacles by impenetrable barriers and induce crystallization at a single site of the medium. For demonstration purposes we have 
tested various types of objects to represent obstacles, including 35~mm Petri dish glued with silicon to a 
bottom of larger Petri dish (Fig.~\ref{expscans}a) and blob of silicon (Fig.~\ref{expscans}b) to make simple obstacles, and 
Blu-Tack$^{\textregistered}$ to make more complex obstacles (Fig.~\ref{expscans}c) and a labyrinth (Fig.~\ref{expscans}d).

\begin{figure}
\centering
\subfigure[]{\includegraphics[width=0.49\textwidth]{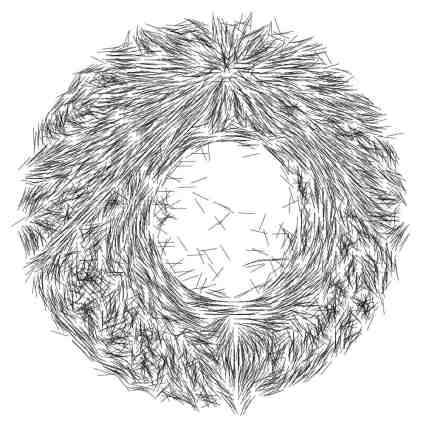}}
\subfigure[]{\includegraphics[width=0.49\textwidth]{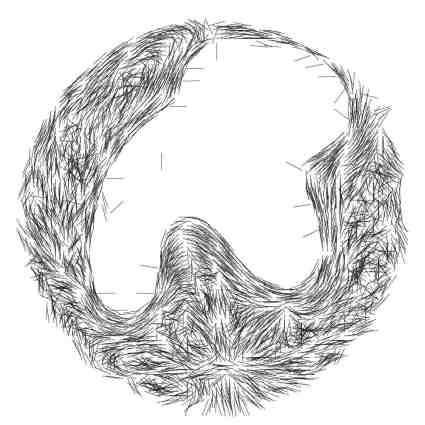}}
\subfigure[]{\includegraphics[width=0.49\textwidth]{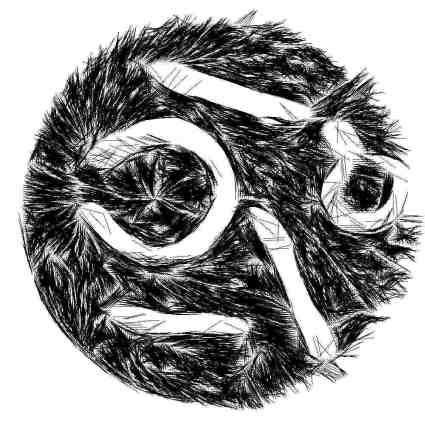}}
\subfigure[]{\includegraphics[width=0.49\textwidth]{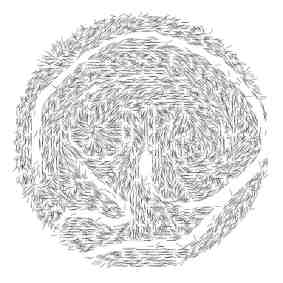}}
\caption{Vectors of crystallization propagation computed from experimental images in Fig.~\ref{expscans}. }
\label{expvectors}
\end{figure}

From the images (Fig.~\ref{expscans}) we calculate a configuration of local vectors (Fig.~\ref{expvectors}). A vector at 
each point indicates the direction from where the wave of crystallization came from. The vectors represent an attracting field, which `pulls' virtual objects, placed in the field, towards the destination (site of initial disturbance). 
We found that the attracting field is approximated by the vectors satisfactory for a wide range of vector parameters: 
vector grid steps 5 (Fig.~\ref{expvectors}c), 10 (Fig.~\ref{expvectors}a) and 20 (Fig.~\ref{expvectors}bd), 
and vector length 20 (Fig.~\ref{expvectors}d), 20 (Fig.~\ref{expvectors}ab) and 50 (Fig.~\ref{expvectors}c).

\begin{figure}
\centering
\subfigure[]{\includegraphics[width=0.49\textwidth]{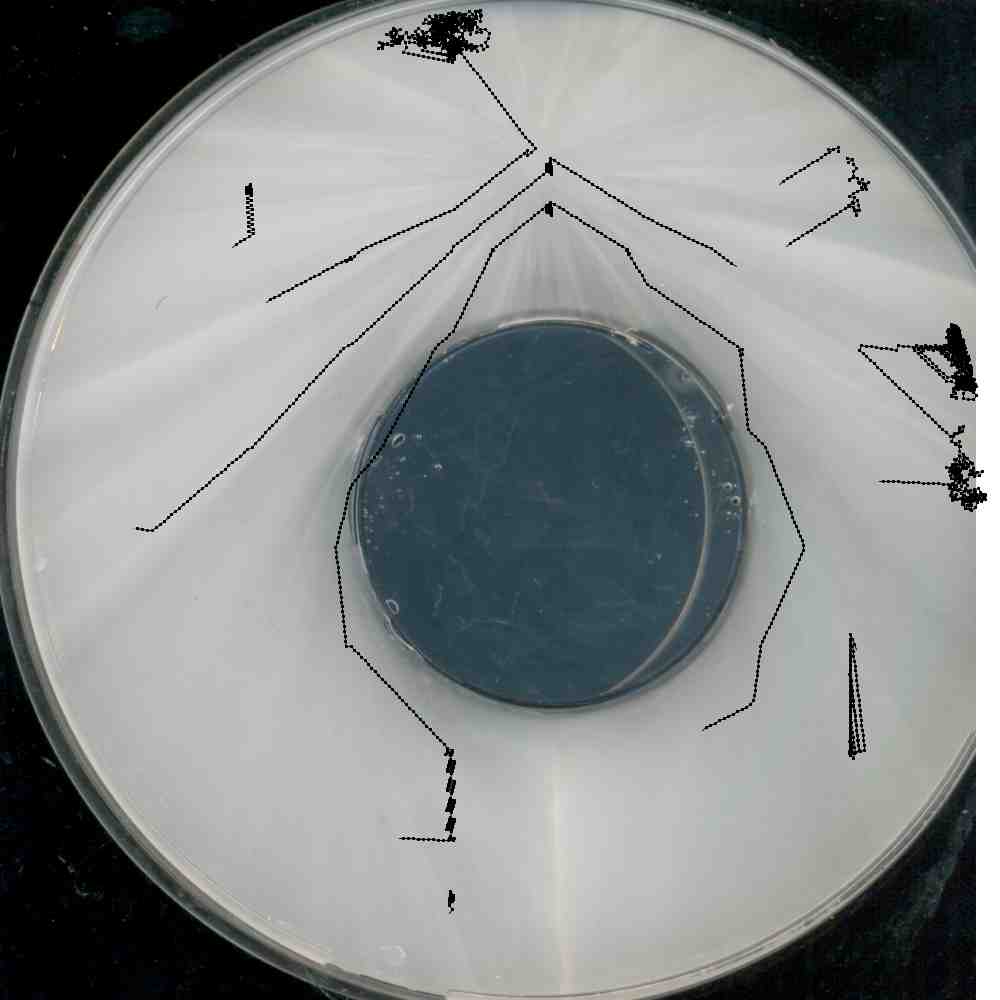}}
\subfigure[]{\includegraphics[width=0.49\textwidth]{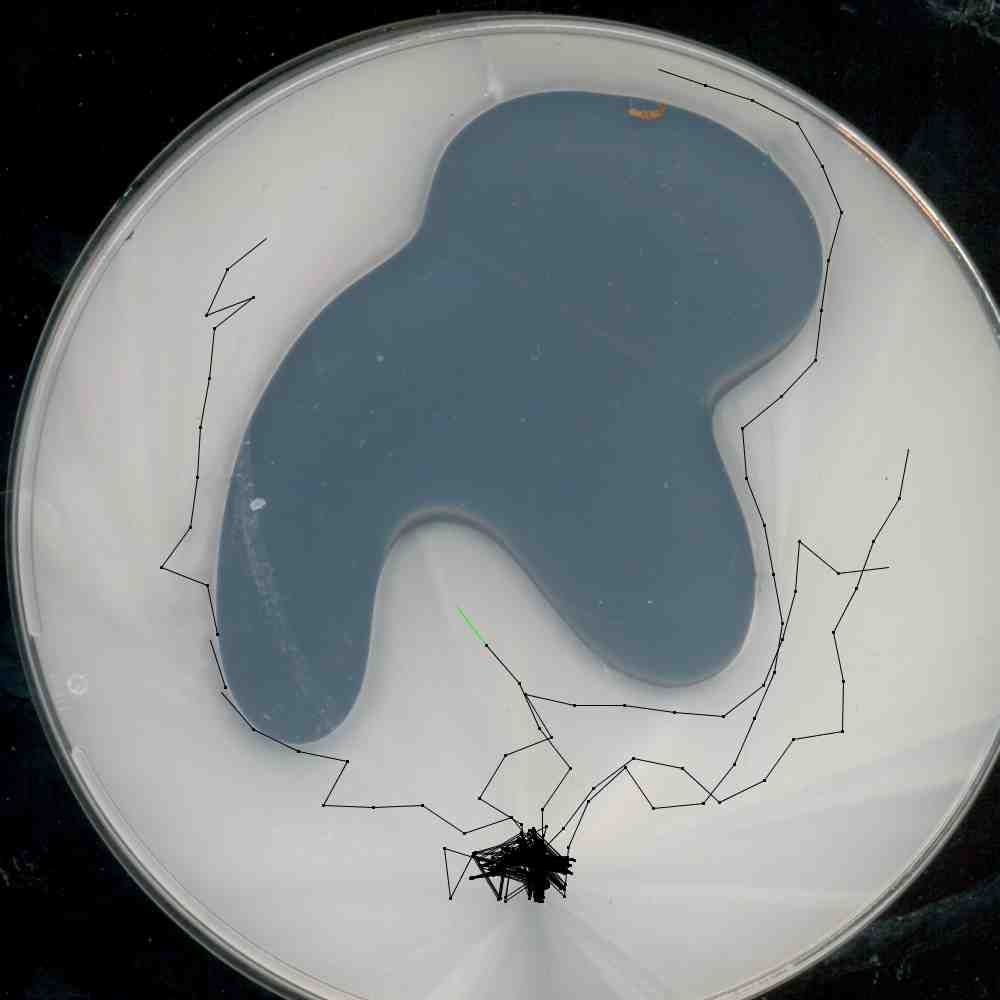}}
\subfigure[]{\includegraphics[width=0.49\textwidth]{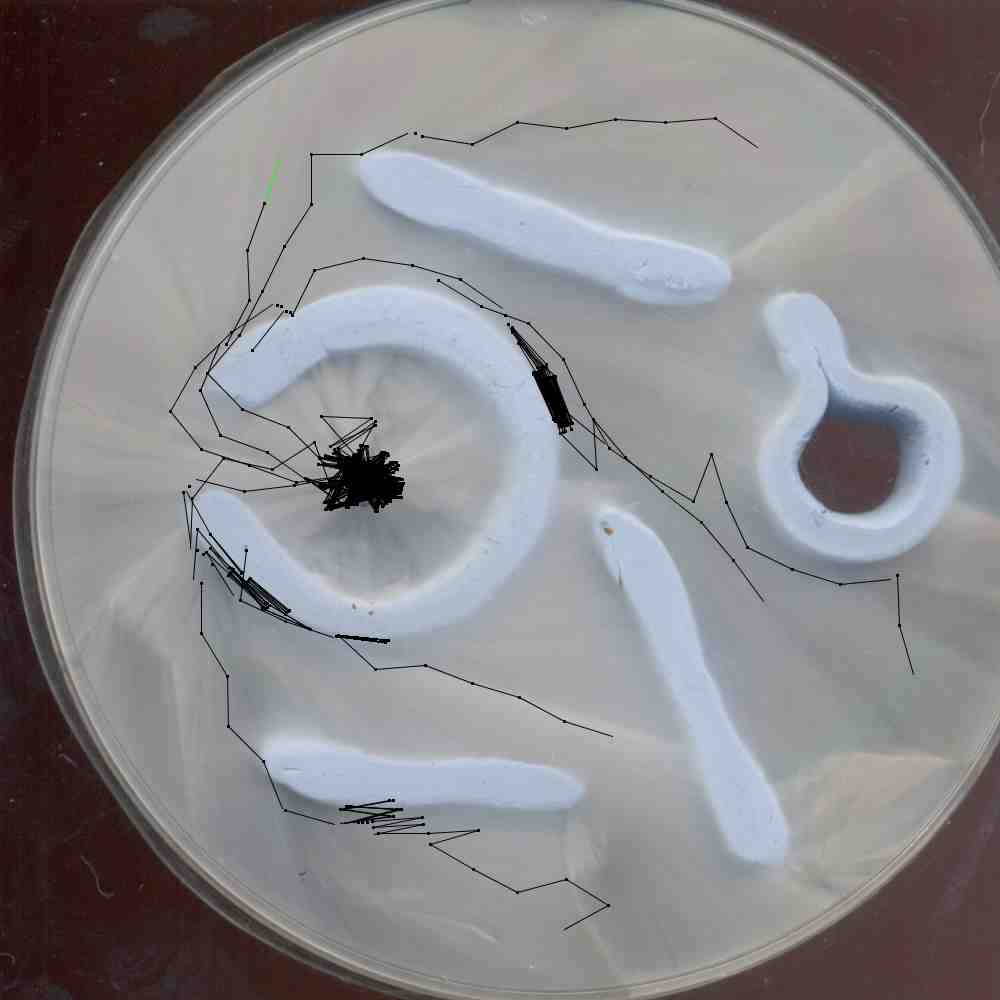}}
\subfigure[]{\includegraphics[width=0.49\textwidth]{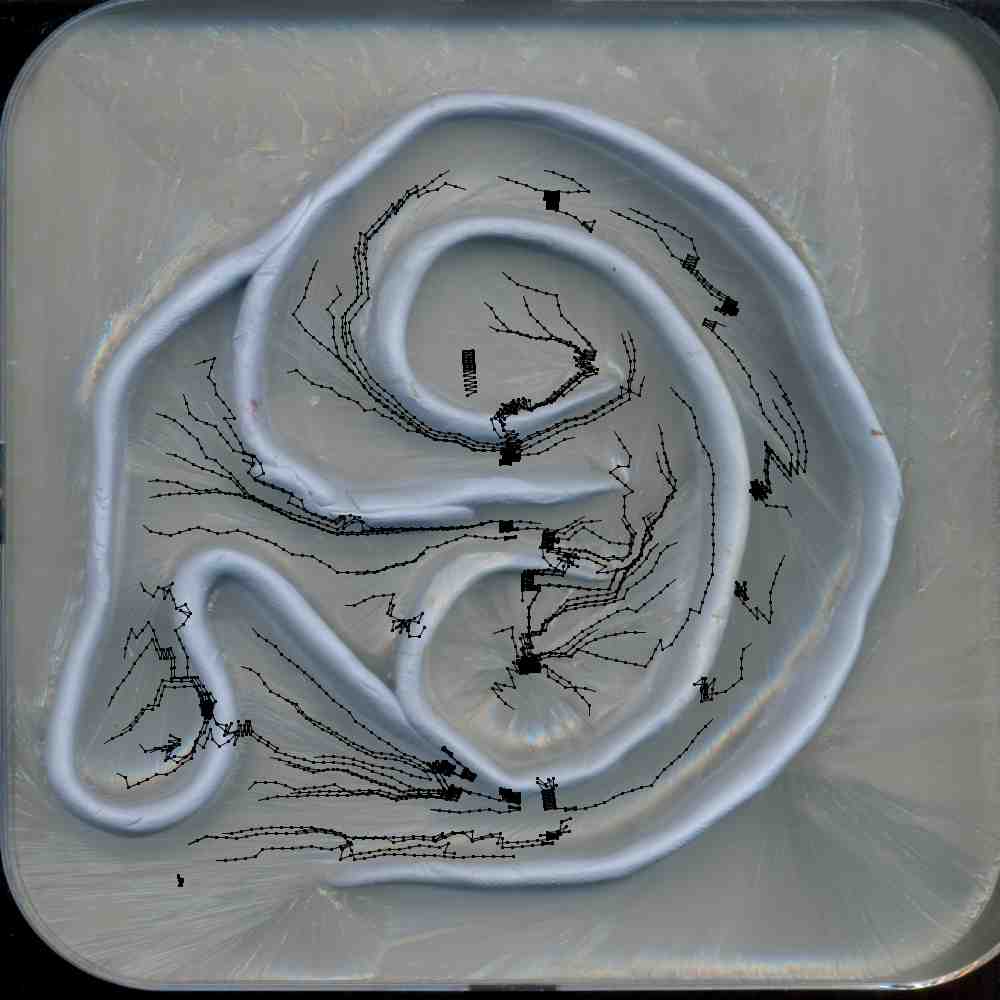}}
\caption{Experimental trajectories of shortest collision-free paths.}
\label{experimental-path}
\end{figure}

To check whether vector configurations extracted from scanned images of crystallized sodium acetate provide sufficient 
representation of many-sources-one-destination shortest paths we employed pixelbots~\cite{adamatzky_2005}. A pixelbot is 
a virtual pixel-sized automaton. The pixelbot moves along the direction of nearest non-zero vector. The pixelbot imitates
a tiny (2-5~mm) in diameter robot which travels along sodium acetate crystals or in the longitudinal grooves between the 
crystals. If there are no non-zero vectors in the pixelbot's neighborhood the pixelbot moves at random~\cite{adamatzky_2005}.

Examples of test-run trajectories of pixelbots traveling in the vector configurations are superimposed on scanned images of 
crystallized sodium acetate in Fig.~\ref{experimental-path}. Pixelbots easily reach the destination in the case of simple 
obstacles (Fig.~\ref{experimental-path}ab). They spent more time wandering at random in arenas with complex obstacles (Fig.~\ref{experimental-path}c). Pixelbots do not perform well in a maze (Fig.~\ref{experimental-path}d): in some parts of
the maze pixelbots navigate satisfactory and travel for substantial distances (see trajectories in the western half of image 
Fig.~\ref{experimental-path}d), in other parts they become stuck forever falling into indefinite loops of repetitive motion
(eastern half of image Fig.~\ref{experimental-path}d).  
  
\section{Implementation of {\sc and} and {\sc or} logical gates}
\label{gates}

We implement logical gates in a geometrically constrained, T-shaped, 
containers with sodium acetate.  Shoulders of a T-shaped container represent 
inputs $x$ and $y$ and vertical stem --- output $z$. When we induce crystallization 
in any one or both shoulders the whole solution in the T-shaped container, 
including $z$-output, becomes crystallized. Assuming liquid phase represents logical {\sc True} and solid phase logical {\sc False}, 
sodium acetate in $T$-shaped container implements {\sc or} gate.

\begin{figure}
\centering
\subfigure[]{\includegraphics[width=0.49\textwidth]{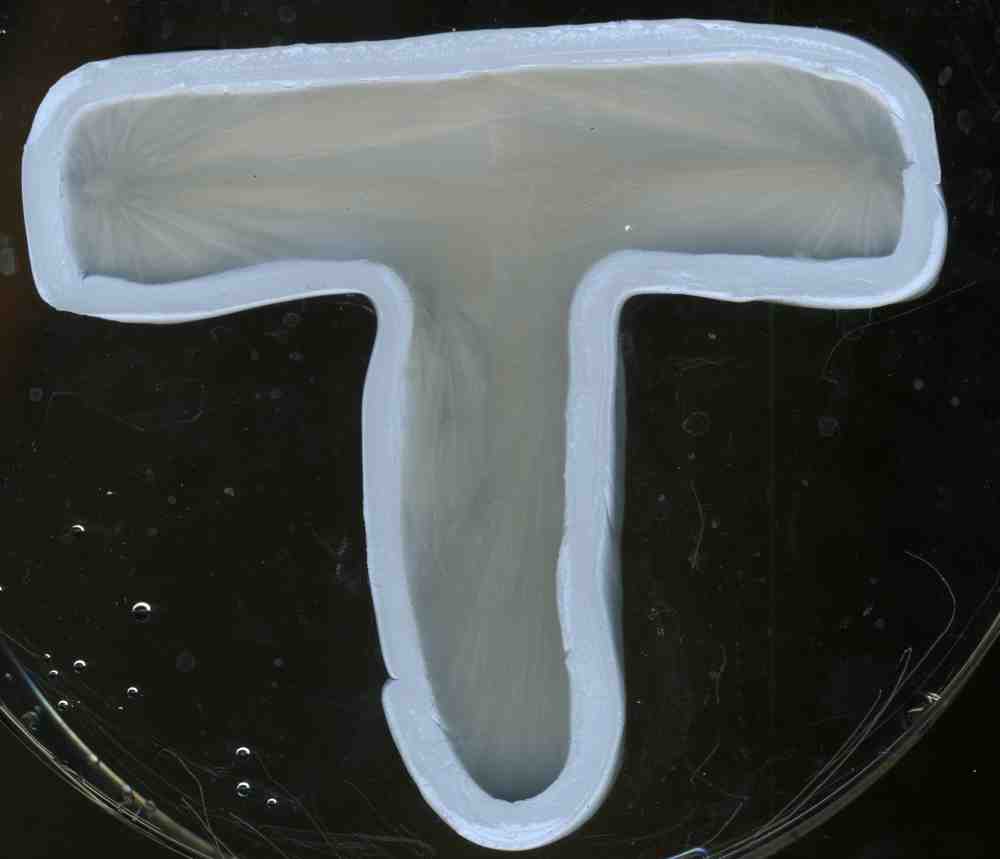}}
\subfigure[]{\includegraphics[width=0.49\textwidth]{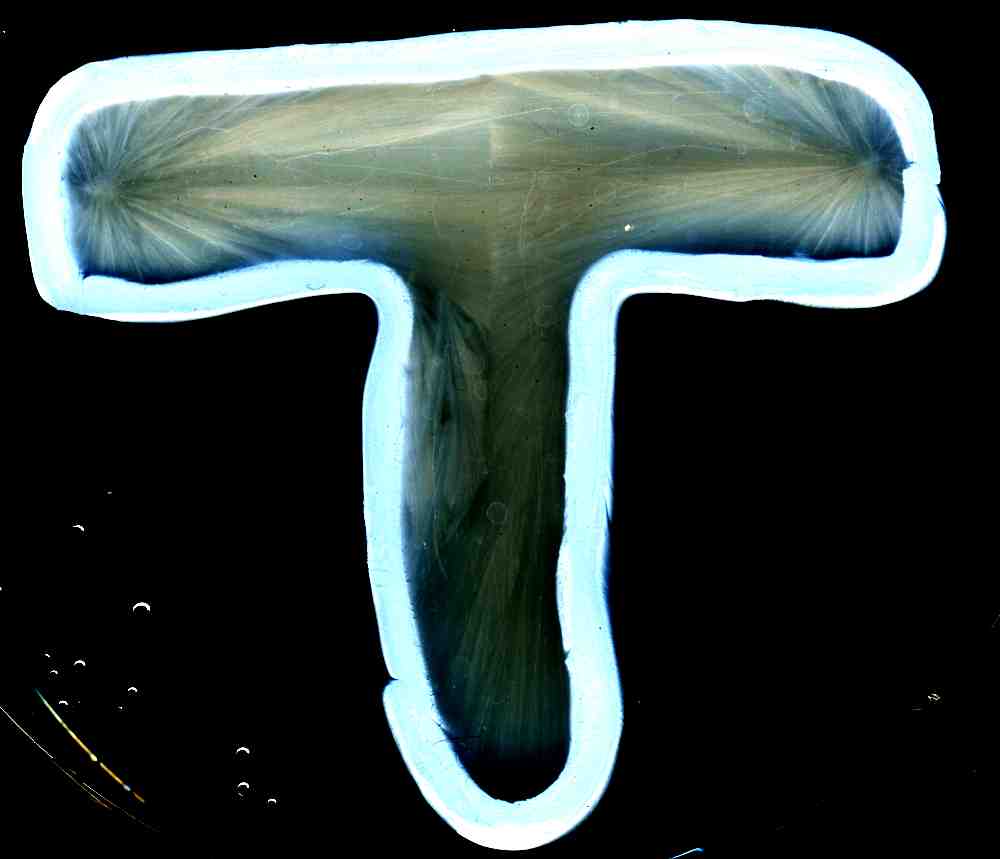}}
\subfigure[]{\includegraphics[width=0.49\textwidth]{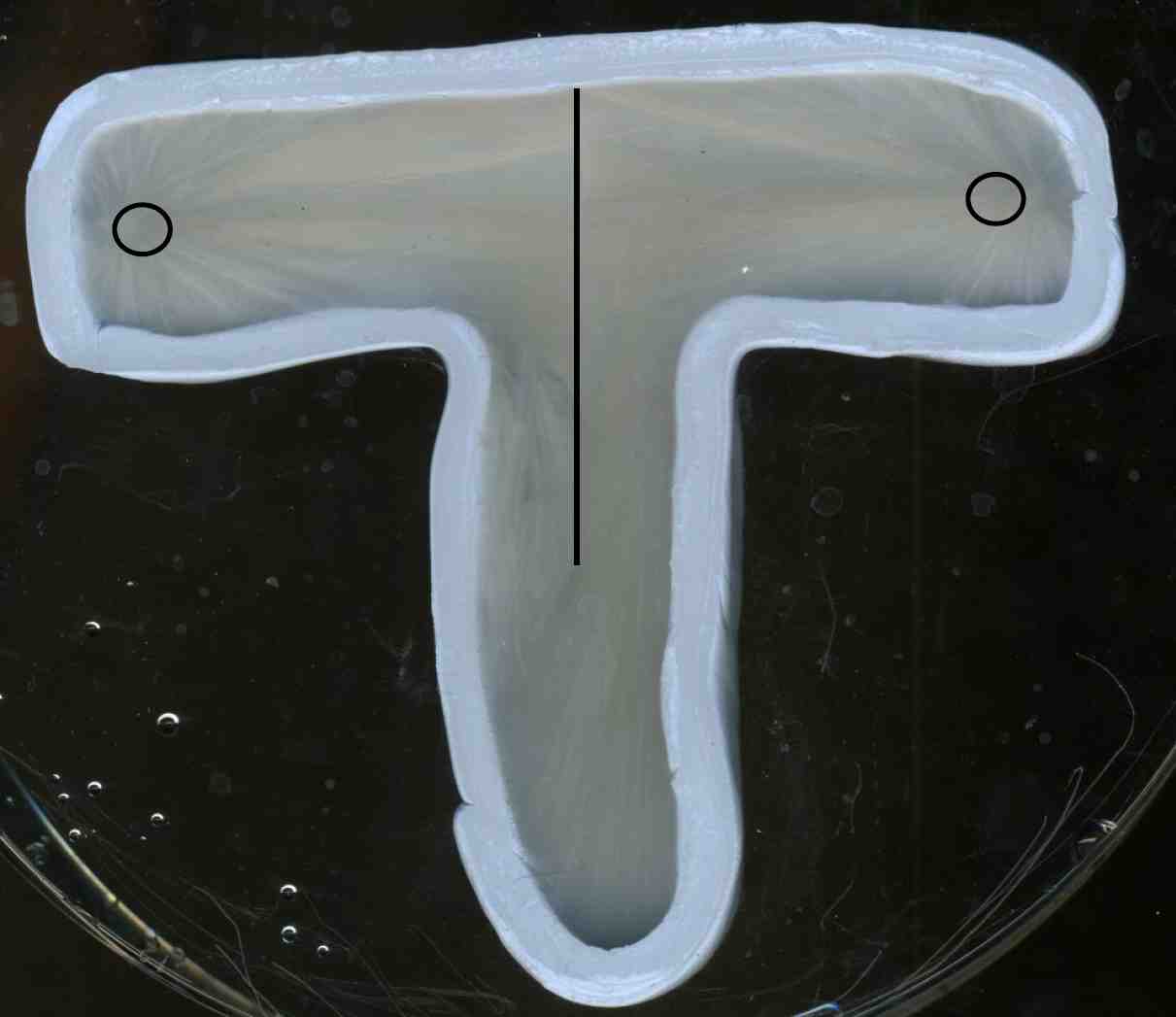}}
\caption{Experimental implementation of {\sc and} gate in sodium acetate: (a)~original scanned image of 
the crystallized sodium acetate, (b)~contrast enhanced image, (c)~scheme of the computation, two sites of crystallization 
induction (corresponding to {\sc True} values of input variables $x$ and $y$) are encircled, bisector separating $x$ and $y$ 
is shown by a line.}
\label{expgate}
\end{figure}

If we induce crystallization in both shoulders of the container crystallization waves collide at 
a junction and form a boundary in result of the collision (Fig.~\ref{expgate}).  
If crystallization is induced only in one of the shoulders no boundary is formed in $z$-output.
This hints to the following interpretation. Presence/absence of crystals in shoulders $x$ and $y$ represent
{\sc True}/{\sc False}.   Presence/absence of boundary (bisector of crystallization induction sites in $x$ and $y$) 
represents  $z=${\sc Truth}/{\sc False}. In such interpretation sodium acetate in $T$-shaped container implements 
{\sc and} gate.

\section{Discussion}
\label{discussion}

We experimentally demonstrated that crystals growing in supersaturated solution of sodium acetate (`hot ice') compute
planar Voronoi diagram, approximate set of collision-free shortest paths in a space with obstacles and 
implement Boolean conjunction and disjunction in a geometrically constrained medium.

Every micro-volume of the sodium acetate solution can input external stimuli/data, participate in crystal growth and 
represent a boundary between collided waves of crystal growth. Data are inputted in the sodium acetate solution 
in parallel, e.g. we use a set of pins to input planar point set to compute a Voronoi diagram. 
Results of computation are also read in parallel, e.g. using optical recording. A micro-volume of the solution can be 
removed or added prior to computation, this is the sign of fault-tolerance. Therefore supersaturated solution of sodium acetate 
is a massively-parallel fault-tolerant computer.

With regards to Voronoi diagram computation the hot ice computer is a highly accelerated version 
of precipitating reaction-diffusion chemical Voronoi processors ~\cite{tolmachiev,ben_2003a,ben_2003b,ben_2004,ben_2009}. 
We demonstrate in experiments that boundaries between crystal patterns perfectly represent edges of Voronoi diagram 
constructed on sites of crystallization induction. We did not show however how exactly the edges can be extracted from 
still images of the crystallized sodium acetate.  This may be a topic of further studies.

In laboratory experiments we proved that hot ice computer approximates a set of many-source-one-destination collision-free shortest paths in a space with obstacles. The shortest path problem is a typical benchmark task for unconventional computing 
devices. There experimental evidences that shortest path can be computed in excitable chemical medium (Belousov-Zhabotinsky reaction)~\cite{agladze,steinbock_1995}, in precipitating chemical medium interfaced with cellular automaton~\cite{adamatzky_delacycostello_2003}, and in plasmodium of \emph{Physarum polycephalum}~\cite{nakagaki_2001a}.

These implementations, although very appealing, are however far from perfect. Some of them require substantial assistance of external
hardware/software in the detection of the `intersection' of  waves initiated in source and destination sites~\cite{agladze,steinbock_1995} or extraction of a shortest path~\cite{adamatzky_delacycostello_2003}, others just do not produce a directed path at all~\cite{nakagaki_2001a}. Hot ice computer physically represents a many-source-one-destination set of shortest paths, where destination is a site of crystallization induction.  If there would be a mobile robot compatible in size with width of crystals, such a robot would find the destination from any site of crystal pattern just by following the orientation  of crystals.
However, our experimental implementation is not ideal: it is sometimes possible to take a wrong route or fall into cyclic route 
by following individual crystals. We will perfect the realization in future.     
  
We demonstrated that two logical gates -- {\sc and} and {\sc or} -- are implemented in crystallization of 
sodium acetate solution. This construction complements experimental realization of {\sc xor} gate in precipitating 
reaction-diffusion chemical medium~\cite{adamatzky_delacycostello_2002}.  One drawback of our approach is that different properties of the medium represent values of output Boolean variables: liquid and solid phases in {\sc or} gate, and presence/absence of a bisector in {\sc and} gate. Thus, in principle, {\sc and} and {\sc or} gates cannot be cascaded. One can consider three valued variables, where liquid phase represents $0$, crystal phase -- $c$ and bisector (a boundary between collided waves of crystallization) -- $b$. Then a T-shaped container with sodium acetate implements the following binary 
composition $\star$: 
$$
\begin{array}{c|ccc}
\star & 0 & b & c \\ \hline
0				& 0 &	c	&	c	\\ 
b				&	c	&	c	&	c	\\
c				&	c	&	c	&	b	\\
\end{array}
$$  
Whether such operation is useful for implementation of logical circuits or not will be studied elsewhere.

Finally we will say a few words about re-usability.  In \cite{rdc} we provide comparative analysis of existing non-linear medium 
computers in terms of re-usability. 

All excitable processors, including Belousov-Zhabotinsky medium, are re-usable chemical computers, because when development of excitation patterns is finished (and a dissipative structure of excitations representing 
result of computation recorded) the medium can be forced back to its resting state, thus being ready for next cycle of computation. However, excitable processors are memoryless, because they do not, as a rule, store results of computation in a non-volatile memory.  

Precipitating chemical processors, e.g.~\cite{adamatzky_delacycostello_2002}, have memory. They represent results of computation in the spatial distribution of precipitate which, when dried, can stay unchanged for years. However, the precipitating chemical processors cannot be re-used. They are not re-usable but ``disposable''.

Hot ice computer is an experimental example of re-usable non-linear medium processor with non-volatile memory. 
Patterns of crystallization, which represent results of computation, stay almost intact for weeks and months. 
As soon as they are heated above 58$^o$C the patterns are transformed into liquid phase, thus being ready for 
next cycle of computation.

\section*{Acknowledgment}

I am indebted to Sebastian Adamatzky for pestering me into playing with the hot ice, and for participating in all experiments, and
to Jeff Jones for editing text and participating in discussions.


\begin{thebibliography}{99}

\bibitem{rdc}
Adamatzky A., De Lacy Costello B., and Asai T., Reaction-Diffusion Computers. Elsevier, 2005.


\bibitem{adamatzky_1994}
Adamatzky~A. Reaction-diffusion algorithm for constructing discrete generalised Voronoi diagram. 
Neural Network World 9 (1994) 6635--6643.


\bibitem{adamatzky_1996}
Adamatzky A. Computation of shortest path in cellular automata. 
Mathematical and Computer Modelling 23 (1996) 105--111.

\bibitem{adamatzky_delacycostello_2002}
Adamatzky~A. and De Lacy Costello~B. 
Experimental logical gates in reaction-diffusion medium: The {\sc xor} gate and beyond.
Physical Review E 66 (2002) 046112.

\bibitem{adamatzky_delacycostello_2003}
Adamatzky~A., De Lacy Costello~B., Melhuish~C. and Ratcliffe~N. 
Experimental reaction-diffusion chemical processors for robot path planning. 
J of Intelligent \& Robotic Systems 37 (2003) 233--249.

\bibitem{adamatzky_2005}
Adamatzky~A., De~Lacy~Costello~B., Skachek~S., Melhuish~C. 
Manipulating objects with chemical waves: Open loop case of 
experimental Belousov–Zhabotinsky medium coupled with simulated 
actuator array. 
Physics Letters A 350 (2005) 201--209. 

\bibitem{adamatzky_ppl_2007}
Adamatzky A. 
Physarum machine: implementation of a Kolmogorov-Uspensky machine on a biological substrate. 
Parallel Processing Letters 17 (2007) 455--467.

\bibitem{demos}
Adamatzky~A. Videos of experiments with hot ice computer. 
\url{http://uncomp.uwe.ac.uk/adamatzky/hot-ice/}.

\bibitem{agladze}
Agladze~K., Magome~N., Aliev~R., Yamaguchi~T., and Yoshikawa~K.
Finding the optimal path with the aid of chemical wave. 
Physica D 106 (1997) 247--254.   

\bibitem{astrov}
Astrov~Y.
Gas-discharge planar semiconductor structures 
as devices for unconventional computing.
Int J Unconventional Computing (2009), in press.

\bibitem{uc_2007}
Adamatzky~A., Bull~L., De~Lacy~Costello~B., Stepney~S., Teuscher~C. 
Unconventional Computing 2007. Luniver Press, 2007.

\bibitem{adamatzky_delacycostello_shirakawa_2008}
Adamatzky~A., De~Lacy~Costello~B., Shirakawa~T. 
Universal computation with limited resources: 
Belousov-Zhabotinsky and Physarum computers. 
Int. J. Bifurcation and Chaos (2008), in press.

\bibitem{adamatzky-bz-trees}
Adamatzky~A.
If BZ medium did spanning trees these would be the same trees as {\emph Physarum} built.
Physics Letters A 373 (2009) 952--956.




\bibitem{uc08}
Calude~C.~S., Costa~J.~F.~G., Freund~R., Oswald~M.
Unconventional Computation: 7th International Conference, UC 2008, Vienna, Austria, 
August 25-28, 2008. Springer, 2008.


\bibitem{uc09}
Calude~C.~S., Costa~J.~F.~G., Dershowitz~N., Freire~E., Rozenberg~G.
Unconventional Computation: 8th International Conference, UC 2009, Ponta Delgada, Portugal, September 7-11, 2009, 
Springer, 2009.

\bibitem{ben_2003a}
De Lacy Costello~B.~P.~J.
Constructive chemical processors --- Experimental evidence that shows this class of programmable pattern 
forming reactions exist at the edge of a highly nonlinear region.
Int J Bifurcation and Chaos 13 (2003) 1561--1564.

\bibitem{ben_2003b}
De Lacy Costello~B.~P.~J. and Adamatzky~A.
On multitasking in parallel chemical processors: experimental results.
Int J Bifurcation and Chaos 
13 (2003) 521--533.


\bibitem{ben_2004}
De Lacy Costello~B., Hantz~P., Ratcliffe~N. 
Voronoi diagrams generated by regressing edges of precipitation fronts
J. Chem. Phys. 120 (2004) 2413.  


\bibitem{ben_2009}
De Lacy Costello~B.~P.~J., Jahan~I., Adamatzky~A., Ratcliffe~N.~M.
Chemical tesselations.
Int. J Bifurcation and Chaos
19 (2009) 619--622.

\bibitem{everitt}
Everitt~M.~S., Jones~M.~L., Kendon~V., Lovett~N.~B., Wagner~R.~C.
Analogue computation with microwaves. Int J Unconventional Computing (2009), in press.

\bibitem{fortune_1986}
Fortune~S. 
A sweepline algorithm for Voronoi diagrams. 
Proc. of the 2nd Annual Symp. on Computational Geometry. Yorktown Heights, New York, 1986, 313–322.   

\bibitem{harding}
Harding~S., Miller~J.~F. and Rietman~E.~A.
Evolution in Materio: Exploiting the physics of materials for computation.
Int J of Unconventional Computing 4 (2008) 155--194. 


\bibitem{klein_1990}
Klein~R.
Concrete and Abstract Voronoi Diagrams. Springer-Verlag, 1985.


\bibitem{mills}
Mills~J.
The nature of the Extended Analog Computer.
Physica D 237 (2008) 1235--1256.


\bibitem{nakagaki_2001a}
Nakagaki T., Yamada H., and Toth A.,
 Path finding by tube morphogenesis in an amoeboid organism. 
Biophysical Chemistry 92 (2001) 47–-52.

\bibitem{nakagaki_iima_2007}
Nakagaki~T., Iima~M., Ueda~T., Nishiura~y., Saigusa~T., Tero~A., Kobayashi~R., Showalter~K.
Minimum-risk path finding by an adaptive amoeba network.
Physical Review Letters 99 (2007) 068104.


\bibitem{reyes}
Reyes~D.~R., Ghanem~M.~G., George~M. 
Glow discharge in microfluidic chips for visible analog computing.
Lab on a Chip 1 (2002) 113--116. 


\bibitem{shamos_preparata}
Prepara~F. and Shamos~M.
Computational Geometry. 
Springer, 1990.

\bibitem{steinbock_1995}
Steinbock~O., T\'{o}th~A., Showalter~K. 
Navigating complex labyrinths: Optimal paths from chemical
waves, Science, 267 (1995) 868-–871.

\bibitem{tolmachiev}
Tolmachiev~D. and Adamatzky~A. 
Chemical processor for computation of voronoi diagram.
Advanced Materials for Optics and Electronics
6 (1996) 191--196.


\bibitem{erokhin}
Erokhin~V.
Organic memristor and bio-inspired information processing. 
Int J Unconventional Computing (2009), in press.

\end{thebibliography}
\end{document}